\documentstyle[psfig,aps,prl,amssymb]{revtex}

\begin{document}
\title{Hall Normalization Constants for the Bures Volumes of 
the $n$-State Quantum Systems}
\author{Paul B. Slater}
\address{ISBER, University of California, Santa Barbara, CA 93106-2150\\
e-mail: slater@itp.ucsb.edu,
FAX: (805) 893-7995}

\date{\today}

\draft
\maketitle
\vskip -0.1cm

\begin{abstract}
We report the results of certain integrations of quantum-theoretic interest,
relying, in this regard, upon recently developed parameterizations of Boya
{\it et al} of the $n \times n$
density matrices, in terms of squared components of
the unit $(n-1)$-sphere and the $n \times n$ unitary matrices. Firstly, we
express the {\it normalized} volume elements of the
Bures (minimal monotone) metric for $n=2$ 
and 3, obtaining thereby  ``Bures prior probability distributions'' over the
two- and three-state systems.
Then, as a first step in extending
 these results to $n>3$, we determine that 
the ``Hall normalization constant'' ($C_{n}$) for
the {\it marginal} Bures prior probablity distribution
over the $(n-1)$-dimensional
simplex of the $n$
 eigenvalues of the $n \times n$ density matrices is, for $n=4$, equal to
${71680 / \pi^{2}}$. Since we also find that $C_{3} = 35 / \pi$, it
follows that $C_{4}$ is simply
equal to  $2^{11} C_{3} / \pi$. ($C_{2}$ itself is known to equal
$2 / \pi$.) The constant $C_{5}$ is also found. It too is associated
with a remarkably simple decompositon, involving the product of the eight
{\it consecutive} prime numbers from 3 to 23.  
We also preliminarily investigate several cases $n>5$, with the use of
quasi-Monte Carlo integration.
 We hope that the various analyses reported will prove
useful in deriving a general formula (which evidence suggests will involve 
the Bernoulli numbers) for the Hall normalization constant
for arbitrary $n$. This would have diverse applications, including quantum
inference and universal quantum coding.

\end{abstract}


\vspace{.1cm}

\tableofcontents

\section{INTRODUCTION}

We make use of recently proposed parameterizations 
\cite{boya} (cf. \cite{zhsl}) --- in terms of squared components of the unit
$(n-1)$-sphere and the $n \times n$ unitary matrices --- of the $n \times n$
density matrices. First (in sec.~\ref{buressection}), we derive (prior)
 probability
distributions of particular interest over both the
 three-dimensional convex set of two-state quantum 
systems and the eight-dimensional convex set of three-state 
quantum systems.
These distributions are the normalized volume elements of the corresponding
Bures metrics on these systems. Hall \cite{hall}
 (cf. \cite{hall2,stasz,rieffel}) 
has contended that such  distributions  correspond
to ``minimal-knowledge'' ensembles, that is 
the most {\it random} ensembles of possible states. In particular,
 for the two-dimensional quantum systems, he 
argues that the Bures metric provides such a minimal-knowledge  ensemble, 
since it ``corresponds to the surface of a unit four-ball, i. e., 
to the maximally symmetric space of positive curvature \ldots 
This space is homogeneous and isotropic, and hence the Bures metric does 
not distinguish a preferred location or direction in the space of 
density operators'' \cite{hall}. Somewhat contrarily  though,
Slater \cite{slaterpla1} has reported results (based on the concept
of comparative noninformativities of priors, first expounded in
\cite{clarke})  that 
indicate  the Bures metric generates ensembles that are less 
noninformative than other (monotone) metrics of interest.

The Bures metric fulfills the role of the
{\it minimal} monotone metric \cite{petzsudar,petzlaa,lesniewski},
 and has been the
focus of a considerable number of studies
 \cite{hall,slaterpla1,hubner1,hubner2,brauncaves,dittmann1,dittmann2}.
``An infinitesimal statistical distance has to be monotone under stochastic
mappings'' \cite[p. 786]{petzsudar}. All stochastically monotone Riemannian
metrics are characterized by means of operator monotone functions. Among
all 
(suitably normalized) operator montone functions $f(t)$ with $f(1)=1$ and $f(t) = t f(t^{-1)}$,
there is a minimal and a maximal one \cite{kubo}.
(The concept of a minimal metric was apparently introduced by Zolotarev in
his extensive paper, ``Metric distances in spaces of random variables
and of their distributions'' \cite[sec. 1.4]{zolotarev}, but it is not 
entirely clear 
that the meaning there is the same as in the terminology ``minimal monotone
metric''.)

We should bear in mind, though --- as emphasized by Petz and
Sud\'ar \cite{petzsudar} --- that, in strong contrast to the classical
situation, in the quantum domain there is not a unique monotone metric, but
rather a (nondenumerable) multiplicity of them.
Their comparative properties need to be evaluated, before deciding which
specific one to employ for a particular application.
We repeat the concluding remarks of Petz and Sud\'ar:
``Therefore, more than one privileged metric shows up in quantum mechanics.
The exact clarification of this point requires and is worth further studies.''

We have previously reported \cite{slaterjmp}
 (in terms of parameterizations other than that
of Boya {\it et al} \cite{boya}) the Bures  probability distribution
for the two-state systems, and also \cite{slaterjpa} for an imbedding of these
systems into a four-dimensional convex set of three-state systems, but
the result below (\ref{new}) for the full eight-dimensional convex set of
three-state systems is clearly novel in nature. In fact, in
 \cite[sec. II.E]{slatertherm2}
we discussed certain (unsuccessful) efforts in these directions
(although the volume element of the {\it maximal} monotone metric --- which
is
not strictly normalizable --- proved more amenable to analysis there).

In sec.~\ref{bureshallsection}, we determine certain necessary elements
for extending the work reported in sec.~\ref{buressection} to the
higher-dimensional quantum systems ($n >3$).
This involves finding the  normalization constant ($C_{n}$), explicitly first
discussed by Hall
\cite[eq. (25)]{hall}, for the {\it marginal} Bures prior probability
distribution over the $(n-1)$-dimensional simplex of the $n$ eigenvalues
of the $n \times n$ density matrices. These constants are found to exhibit
quite remarkable number-theoretic properties. It would, therefore, certainly
be of substantial interest to find a general formula for $C_{n}$.
Knowledge of the value of $C_{n}$, together with that of the invariant
Haar element for $SU(n)$ --- apparently presently available, however, in
suitably parameterized form (cf. \cite{Zycz1,Zycz2}) for
$n \leq 3$ \cite{byrd,byrdsudarshan} --- would allow
 one to construct the Bures prior
probability distribution itself for the $n$-level quantum systems.

In an extensive study \cite{kratt}, Krattenthaler and Slater
examined (in the
framework of the {\it two}-state systems) the hypothesis that the normalized
volume element of the Bures metric would function in the quantum domain in
a role parallel to that fulfilled classically by the ``Jeffreys'
 prior'' --- that is, the normalized volume element of the unique monotone/Fisher
information metric \cite{kass,kwek}.
In particular, they were interested in \cite{kratt}
 in the possibility of extending 
certain (classical) seminal results of Clarke and Barron \cite{cb1,cb2}.
They  did conclude, however, contrary to their working hypothesis, that 
the normalized volume element of the Bures metric does not in fact strictly
fullfill the same role as the Jeffreys' prior (in yielding both the
asymptotic minimax and maximin redundancies for universal
coding/data compression), but it appears to come
remarkably close to doing so
(cf. Fig.~\ref{comparison}).
In sec.~\ref{QUASI}, for the cases $n=2$ and 3, the 
``quasi-Bures'' prior probability distributions are presented 
that appear to fulfill
this distinguished information-theoretic role.

\section{BURES PROBABILITY DISTRIBUTIONS OVER THE ${n \times n}$ DENSITY
MATRICES} \label{buressection}

Boya {\it et al} \cite{boya} have recently ``shown that the mixed state
density matrices for $n$-state systems can be parameterized in terms of
squared components of an $(n-1)$-sphere and unitary matrices''.
The mixed state density matrix ($\rho$) is represented in the form,
\begin{equation} \label{one}
 \rho = U D U^{\dagger} ,
\end{equation}
where $U$ denotes an $SU(n)$ matrix, $U^{\dagger}$ its conjugate
transpose and $D$ a diagonal density operator,
the diagonal entries ($d_{i}$'s)
 of which --- being the eigenvalues of $\rho$ --- are the squared
 components of the
$(n-1)$-sphere. Thus, for $n =2$,
\begin{equation} \label{two}
D = \pmatrix{\cos^{2}{\theta /2} & 0 \cr 0 & \sin^{2}{\theta /2} \cr} \qquad
0 \leq \theta \leq {\pi \over 2},
\end{equation}
and, for $n = 3$,
\begin{equation} \label{three}
D = \pmatrix{\cos^{2}{\phi /2} \sin^{2}{\theta /2} & 0 & 0 \cr
   0 & \sin^{2}{\phi /2} \sin^{2}{\theta /2} & 0 \cr
   0 & 0 & \cos^{2}{\theta /2} \cr} \qquad 0 \leq \theta, \phi \leq \pi.
\end{equation}
(Note the differences in the ranges of angles used in the two
cases. This will be commented
upon in sec.~\ref{n4bures}.)

Biedenharn and Louck have
 presented \cite[eq. (2.40)]{bied} the parameterization of an element of
$SU(2)$, 
\begin{equation} \label{su2}
U(\alpha \beta \gamma) = e^{-i \alpha \sigma_{3} /2} e^{-i \beta \sigma_{2} /2}
e^{-i \gamma \sigma_{3} /2},
\end{equation}
in terms of the Pauli matrices ($\sigma_{i}$'s) and
three Euler angles --- $ 0 \leq \alpha < 2 \pi,
0 \leq \beta \leq \pi, 0 \leq \gamma < 2 \pi$ --- with an associated
invariant Haar measure \cite[eq. (3.134)]{bied},
\begin{equation} \label{invariant1}
\mbox{d} \Omega_{2} = {1 \over 8} \mbox{d} \alpha  \mbox{d} \gamma 
\sin{\beta} \mbox{d} \beta.
\end{equation}
Byrd \cite{byrd} (cf. \cite{byrdsudarshan}) has extended this approach to 
$SU(3)$. He obtains 
\begin{equation}
U(\alpha,\beta,\gamma,\kappa,a,b,c,\zeta) = e^{i \lambda_{3} \alpha}
e^{i \lambda_{2} \beta} e^{i \lambda_{3} \gamma} e^{i \lambda_{5} \kappa}
e^{i \lambda_{3} a} e^{i \lambda_{2} b} e^{i \lambda_{3} c}
e^{i \lambda_{8} \zeta},
\end{equation}
where  $\lambda_{i}$ denotes one of  the 
eight $3 \times 3$ Gell-Mann matrices \cite{lukach}.
The corresponding invariant element is
\begin{equation} \label{invariant2}
\mbox{d} \Omega_{3} = \sin{2 \beta} \sin{2 b} \sin{2 \kappa} {\sin^{2}{\kappa}}
\mbox{d} \alpha \mbox{d} \beta \mbox{d} \gamma \mbox{d} \kappa \mbox{d} a
\mbox{d} b \mbox{d} c \mbox{d} \zeta,
\end{equation}
with the eight Euler angles having the ranges,
\begin{equation} \label{ranges}
0 \leq \alpha, \gamma, a, c < \pi, \quad 0 \leq \beta, b, \kappa \leq
{\pi /2}, \quad 0 \leq \zeta  < \sqrt{3}.
\end{equation}
For our purposes, the Euler angle $\gamma$ for the case $n=2$ and
 $c$ and $\zeta$ in the case $n=3$ are irrelevant, as they
``drop out''  in the formation of the product
(\ref{one}). (I thank M. Byrd for this important observation.)
 So, we will employ below the appropriate {\it conditional}
 versions
of these invariant measures (\ref{invariant1}) and (\ref{invariant2}) --- the
condition 
(a technical statistical term) 
corresponding, of course, to the ignoring of the indicated angles.

The Bures metric itself is expressible in the form \cite[eq. (10)]{hubner1}
\begin{equation} \label{bures}
d_{B}^{2}(\rho,\rho + \mbox{d} \rho) = \sum_{i,j =1}^{n} {1 \over 2}
{{|<i|\mbox{d} \rho |j>}^{2} \over d_{i} + d_{j}},
\end{equation}
where $|i>$ denotes the eigenvectors of the $n \times n$ density matrix
$\rho$, $<j|$, the corresponding complex conjugate (dual) vectors, and 
the $d$'s are the associated eigenvalues.
The parameterization of Boya {\it et al} is, then, particularly
convenient, since the eigenvalues and eigenvectors of $\rho$ are
immediately available. Our chief  concern must, then, be to compute the
complete
Jacobian of the transformation to the set of parameters 
of Boya {\it et al}.
(We shall note for further reference the occurrence of the term $d_{i} + d_{j}$
in (\ref{bures}). This, is of course, simply proportional to the arithmetic
mean, $(d_{i} +d_{j})/2$. By replacing this term by (twice) the
 {\it exponential/identric} mean (\ref{qi13}) of $d_{i}$ and $d_{j}$,
that is $2 I(d_{i},d_{j})$,
we shall obtain the particular ``quasi-Bures'' distributions described
in sec.~\ref{QUASI}.)

\subsection{The Bures case $n=2$}

For the case $n=2$, the volume element of the Bures  metric 
(\ref{bures}) is proportional to the
product of the inverse of the square root of the determinant of 
$\rho$ (or, equivalently, the determinant of $D$) with two Jacobians. The
 first Jacobian (in line with the familiar practice
in the theory of random matrices \cite[eq. (3.3.5)]{mehta}) is 
itself the product of
$(d_{1}-d_{2})^2$ and the (conditional) invariant element (\ref{invariant1}).
The second Jacobian, ${\sin{\theta} / 2}$,
 corresponds simply to the transformation from cartesian
coordinates to the squared polar coordinates employed in
(\ref{two}). Simplifying and normalizing the full  product, we arrive at
the probability density for the normalized volume element of the Bures
metric over the three-dimensional convex set
of two-state quantum systems. This density is
\begin{equation} \label{n2}
\mbox{d} p_{Bures:2}(\theta,\alpha,\beta) = {\cos^{2}{\theta}
 \sin{\beta} \over {\pi^{2}}} \mbox{d} \theta 
\mbox{d} \alpha \mbox{d} \beta  \qquad 0 \leq \theta \leq {\pi \over 2},
\quad 0 \leq \alpha \leq 2 \pi, \quad 0 \leq \beta \leq \pi.
\end{equation}
The expected values of the eigenvalues are, then, ${1/2 \pm 4 / 3 \pi}$.

\subsection{The Bures case $n=3$} \label{n3bures}

\begin{figure}
\centerline{\psfig{figure=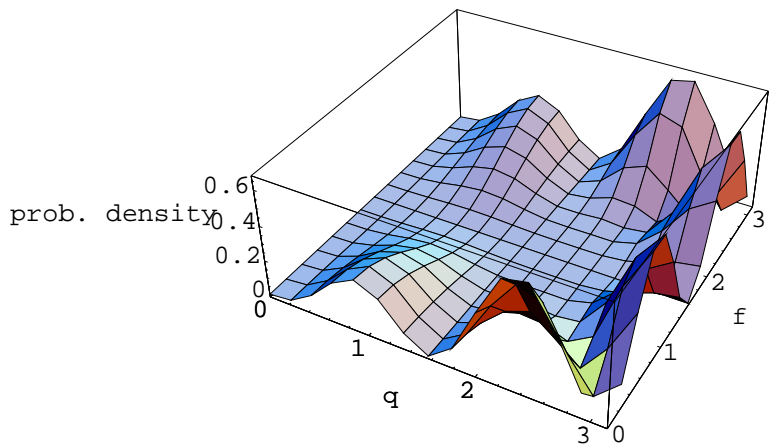}}
\caption{Bivariate marginal Bures prior probability distribution over the
variables
$\theta$ and $\phi$
 for three-state systems}
\label{twodim}
\end{figure}

For the case $n=3$, the volume element of the Bures metric is equal to the
product of: (i)  two Jacobians again, one of which now has the form
$\big((d_{1}- d_{2}) (d_{1}-d_{3}) (d_{2}-d_{3})\big)^2$ multiplied by the
 (conditional)
invariant measure (\ref{invariant2}), while the other, $(\cos{{\theta \over 2}}
\sin^{3}{{\theta \over 2}} \sin{\phi})/2$, corresponds to the
 transformation to squared spherical coordinates used in
(\ref{three}); and (ii)
the reciprocal of the product of the square root of the determinant of $\rho$ 
(or, equivalently, of $D$) and
the difference between the sum of the three principal minors of order two
of $\rho$ (or, equivalently, of $D$)
 and the determinant itself. Since $|\rho|^{1/2} =
(\cos{{\theta \over 2}} \sin^{2}{{\theta \over 2}} \sin{\phi})/2$, it can be
seen that
considerable cancellation occurs between the numerator and the
denominator of the full product.
 The normalization of the resultant volume element
required considerable manipulations using MATHEMATICA
(basically involving reducing the problem to the simplest possible
form at each stage of the integration process). We
obtained the following Bures prior probability density
over the eight-dimensional convex set of three-state (spin-1)
quantum systems,
\begin{equation} \label{new}
\mbox{d} p_{Bures:3}(\theta,\phi,\alpha,\beta,\gamma,\kappa,a,b) =
 {35 u  
 \over 128 {\pi}^4
(35 + 28 \cos{\theta} + \cos{2 \theta} -8 \cos{2 \phi}
 {\sin^{4}{{\theta \over 2}}})} 
\mbox{d} \theta \mbox{d} \phi \mbox{d} \widetilde{\Omega}_{3},
\end{equation}
where
\begin{equation} \label{newvariable}
u   = {\sin^{3}{{\theta \over 2}}}
 \big((35 + 60 \cos{\theta} + 33 \cos{2 \theta} )
\cos{\phi} - {8 \cos{3 \phi}} {{\sin^{4}{{\theta \over 2}}}})^2 \big),
\end{equation}
and the conditional invariant element (cf. (\ref{invariant2})) is 
\begin{equation} \label{conditionalhaar}
 \mbox{d} \widetilde{\Omega}_{3} = \sin{2 \beta} \sin{2 b} \sin{2 \kappa}
{\sin^{2}{\kappa}}
\mbox{d} \alpha 
\mbox{d} \beta \mbox{d} \gamma \mbox{d}
\kappa \mbox{d} a \mbox{d} b.
\end{equation}
The eight variables have the previously indicated ranges ((\ref{three}),
(\ref{ranges})).
In Fig.~\ref{twodim}, we display the two-dimensional marginal probability 
distribution of
(\ref{new}) over the parameters $\theta$ and $\phi$
(which are invariant under unitary transformations of $\rho$).
Let us note that the fully mixed state --- corresponding to the 
$3 \times 3$ diagonal
density matrix with entries equal to $1/3$ --- is obtained at
$\phi = \pi/2, \theta = 2 \cos^{-1}({1/\sqrt{3})} \approx 1.91063$.
The probability density (\ref{new}) is zero  at this distinguished 
point, as well as
along the loci $\theta = 0$ and $\phi = \pi /2$. (Wherever at least two
of the eigenvalues of $\rho$ or, equivalently $D$, are equal, the density
is zero.)

\begin{figure}
\centerline{\psfig{figure=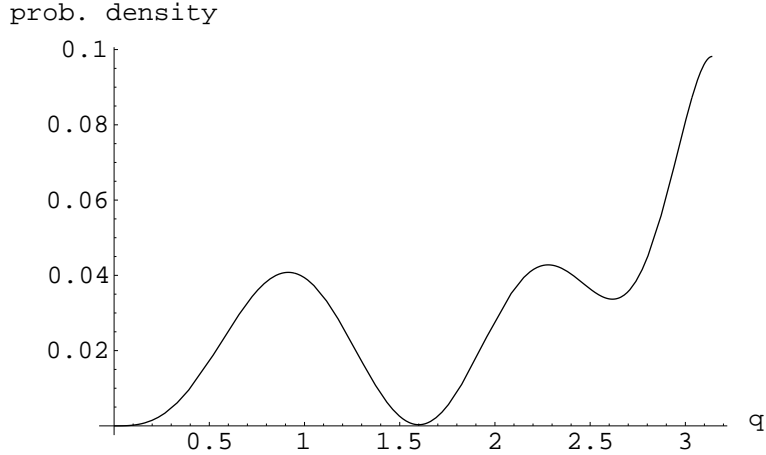}}
\caption{Univariate marginal Bures prior probability distribution over the
variable $\theta$ for
three-state systems}
\label{onedim1}
\end{figure}

The one-dimensional marginal probability density (Fig.~\ref{onedim1}),
obtained by integrating (\ref{new}) over all variables except $\theta$, is
\begin{equation}
\mbox{d} \tilde{p}_{Bures:3}(\theta) =
 {35 \over 256} (-1533 + 2816 \cos{{\theta \over 2}} - 1988 \cos{\theta} +
1152 \cos{{3 \theta \over 2}} - 447 \cos{2 \theta} + 128 \cos{{5 \theta 
\over 2}})
\sin^{3}{{\theta \over 2}} \mbox{d} \theta.
\end{equation}
The relative maxima of this density are located at .914793, 2.2795
and $\pi$, while the relative
minima are at 0, 1.59995 and 2.61732.

\begin{figure}
\centerline{\psfig{figure=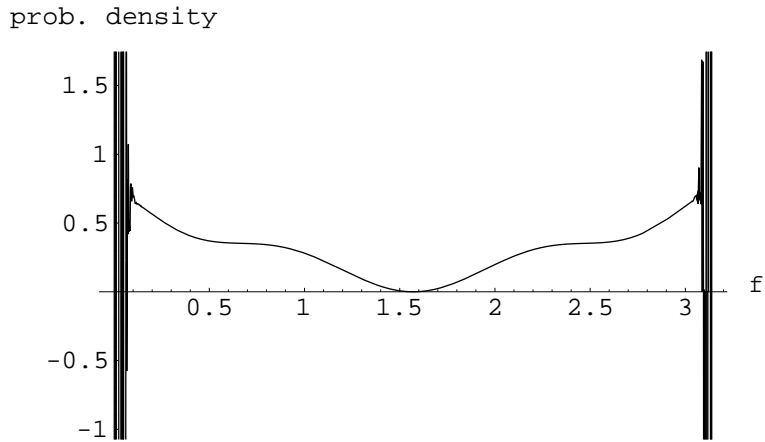}}
\caption{Univariate marginal Bures prior probability distribution over
the variable $\phi$ for three-state
systems}
\label{onedim2}
\end{figure}

The one-dimensional marginal probability density 
(Fig.~\ref{onedim2}), obtained by integrating (\ref{new}) over all
variables except $\phi$,
is
\begin{equation} \label{othermarginal}
\mbox{d} \tilde{\tilde{p}}_{Bures:3}(\phi) =
 {1 \over 768 \pi} \big(\cot{\phi} {\csc^{8}{\phi}}
 \big(110100480 \tan^{-1}{[\cot{\phi \over 2}]} {\cos^{12}{{\phi \over 2}}} - 26880 (792 (2 \pi -\phi) \cos{\phi}
+ 8 \pi
\end{equation}
\begin{displaymath}
 (55 \cos{3 \phi} + 3 \cos{5 \phi})
 + \phi (495 \cos{2 \phi} - 220 \cos{3 \phi} +66 \cos{4 \phi} -12 \cos{5 \phi} +  \cos{6 \phi})) +
16885656 \sin{2 \phi}
\end{displaymath}
\begin{displaymath}
 + 5069937 \sin{4 \phi} + 167012 \sin{6 \phi} -
3 ( 4139520 \phi + 124 \sin{8 \phi} - 4 \sin{10 \phi} + \sin{12 \phi})\big)\big) 
\mbox{d} \phi.
\end{displaymath}
The limits of (\ref{othermarginal}) as $\phi$ approaches 0 and $\pi$ are
both equal to ${20 \over 9 \pi} \approx .707355$. (The density is highly
oscillatory in the vicinity of these boundary points.)
MATHEMATICA does, in fact, perform a {\it symbolic/exact} integration of
(\ref{othermarginal}), yielding the result 1. (Note that for our analyses
below for $n>3$, we have found it
necessary to rely upon {\it numerical} integrations, although the results
obtained do appear to indicate that
exact solutions  exist, which in principle might be found with a powerful
enough computer.)
However, several warning messages are generated en route to this result,
concerning indeterminate expressions and inconsistencies in the arguments
of MeijerG functions (which are very general forms of hypergeometric
functions) \cite{MeijerG}.

\section{HALL NORMALIZATION CONSTANTS FOR {\it MARGINAL} BURES
 PROBABILITY DISTRIBUTIONS OVER THE $(n-1)$-DIMENSIONAL
SIMPLEX OF THE $n$ EIGENVALUES OF THE $n \times n$ DENSITY MATRICES}
\label{bureshallsection}

Let us note that Hall \cite[eq. (24)]{hall} has, in fact, 
 given an explicit formula
for the volume element of the Bures metric on the
$n \times n$ density matrices. This is (converting to the notation used
above),
\begin{equation} \label{hallformula}
\mbox{d} V_{Bures:n} =
 {\mbox{d} d_{1} \ldots \mbox{d} d_{n} \over (d_{1} \ldots d_{n})^{1/2}}
\prod_{i < j}^{n} 4 {(d_{j} -d_{k})^{2} \over d_{i} + d_{j}} 
\mbox{d} x_{ij} \mbox{d} y_{ij},
\end{equation}
where the real part of the $ij$-entry of the diagonalizing unitary
matrix $U$ in formula (\ref{one}) is represented by 
$x_{ij}$ and the imaginary part by $\pm i y_{ij}$.
Since in the parameterization of Boya {\it et al} 
\cite{boya}, which we have employed, one
uses not these $x$'s and $y$'s, but rather the Euler angles
parameterizing the unitary matrix
$U$, we have been compelled to replace the differential elements 
$\mbox{d} x_{ij} \mbox{d} y_{ij}$ in
(\ref{hallformula}) by the corresponding conditional form ($\mbox{d}
 \widetilde{\Omega}$)
of the invariant (Haar) measures (\ref{invariant1}) and (\ref{invariant2}).
 One can, then, confirm that our presentation and results are 
fully consistent with the use of
 (\ref{hallformula}), bearing in mind the unit trace
requirement that $d_{1} + \ldots +d_{n} =1$.
Hall \cite[eq. (25)]{hall} also expressed the marginal
 Bures probability distribution over the space of $n$ eigenvalues of
$\rho$ as
\begin{equation} \label{hallconstant}
 \tilde{p}_{Bures:n}(d_{1},\ldots,d_{n}) =
 C_{n} {\delta (d_{1} + \ldots + d_{n} -1) \over (d_{1} \ldots d_{n})^{1/2}}
 \prod_{i < j}^{n} {(d_{i} -d_{j})^2 \over d_{i}+ d_{j}}.
\end{equation}
We shall report the values of the ``Hall constants'' $C_{n}$ for $n =2, 3, 4$
and 5, immediately below. (The results for $n > 2$ are, apparently, new.)

\subsection{The  Hall constants for $n=2$ and $n=3$} \label{proximate}

If, for consistency with our further results for $n>2$, we take $\theta \in
[0,\pi]$ and not $[0,\pi /2]$ as in \cite{boya}, then we find that
$C_{2}$  equals $2 / \pi
\approx .63662$ (cf. \cite[eq. (30)]{hall}).
From the results of the  analysis in sec.~\ref{n3bures},
 we are able to determine,
for the first time, apparently,  
that $C_{3} = 35 / \pi \approx 11.1408$.
Let us note here that 35 is, of course, simply the product of
{\it proximate} or {\it neighboring} prime numbers, that is,
$35 = 5 \cdot 7$.

\subsection{The Hall constant for $n=4$ and associated
methodology} \label{n4bures}

To continue the full line of research reported here 
for the cases $n=2, 3$, to
$n > 3$, it would be useful to extend
 the work of Byrd and Sudarshan
\cite{byrd,byrdsudarshan} on the Euler angle parameterization of $SU(3)$
to such higher $n$.
However, computation of the Hall constants $C_{n}$ (\ref{hallconstant})
 does not depend on
parameterizations of $SU(n)$.
We have, in fact, been able to obtain 
exceedingly strong {\it numerical} evidence
that $C_{4}$ is, in fact, equal to ${71680 / \pi^{2}} \approx 7262.7$. 
Let us, first, mention some methodological considerations
useful in deriving this result (and, in general, $C_{n}$, $n > 4$).

The parameterizations of Boya {\it et al} \cite{boya}
 of the diagonal $2 \times 2$ and $3 \times 3$
matrices differ, in that in the $2 \times 2$ case (\ref{two}) only matrices in which
the (1,1)-entry is at least as great as the (2,2)-entry are generated
(due to the restriction of the angular parameter, $\theta$, to the
range $[0, \pi /2]$),
while in the $3 \times 3$ case (\ref{three}),
 no order is imposed on the diagonal 
entries (the parameters $\theta$ and $\phi$ both varying freely between
0 and $\pi$). Now, in performing 
(the apparently necessary) numerical (as opposed to symbolic) 
integrations to obtain the Hall
constants $C_{n}$ $(n > 3)$, it seems to be considerably more computationally
effective to integrate over 
only those diagonal matrices in which (say) the (1,1)-entry
is no less than the (2,2)-entry, which in turn is no less than the 
(3,3)-entry, {\it etc.} (This helps to minimize troublesome
oscillations.)
 Then, the result can be multiplied  by the number ($n!$) of
permutations of $n$ objects to yield
$C_{n}$, since the result of the integration must be invariant under
any other of the $n!-1$
possible orderings (permutations) that can be
imposed on the diagonal entries of the
$n \times n$ diagonal matrices. In 
precisely this manner, we were able to obtain
(using  the numerical integration of interpolating function command of 
MATHEMATICA)  the result
$71680.000001/ \pi^{2}$ for $C_{4}$. This we take as overwhelming
evidence that $C_{4}$, in fact, equals $71680 / \pi^{2}$, particularly so,
since 71,680 has the highly structured prime decomposition of 
$2^{11} \cdot 5 \cdot 7$. (It is also interesting to note that $35 = 5 \cdot 7$
appears in the numerator of $C_{3}$, that is, $35 / \pi$, or, to the
same effect, $C_{4} = 2^{11} C_{3}/ \pi$.

To illustrate the procedure followed,
 let us first parameterize the $4 \times 4$ nonnegative diagonal
matrices of trace unity in the following fashion
 (cf. (\ref{two}), (\ref{three})),
\begin{equation}
D = \pmatrix{ \cos^{2}{\theta /  2} & 0 & 0 & 0 \cr
0 & \sin^{2}{\theta / 2} \cos^{2}{\phi / 2} & 0 & 0 \cr
0 & 0 & \sin^{2}{\theta / 2} \sin^{2}{\phi / 2}
 \cos^{2}{\zeta / 2}& 0 \cr
0 & 0 & 0 & \sin^{2}{\theta / 2} \sin^{2}{\phi / 2} \sin^{2}{\zeta
/ 2} \cr}.
\end{equation}
Then, the 
(truncated) region of integration employed above --- corresponding to
successively nonincreasing diagonal entries --- can be described
as $\zeta \in [0, \pi / 2], \phi \in [0, f(\zeta)], \theta \in  [0, f(\phi)]$,
where $f(x) = 2 \cot^{-1}{(\cos{x/2})}$, rather than
$\zeta, \phi, \theta \in [0, \pi]$, as in the apparently suggested
parameterization of Boya {\it et al}
\cite{boya}, which would yield all possible diagonal matrices, without
regard to the ordering of their elements.

If we construct a similar truncated region of integration in the case $n=3$,
 then we find that the expected values of the eigenvalues are
.802393, .181878 and .0157299.
In the $n=2$ case, the analogous values are (as previously noted),
$1/2 \pm 4 / 3 \pi$, that is  .924413, and .0755868.

\subsection{The Hall constant for $n=5$} 

 We have also attempted to
compute $C_{5}$, in the manner of sec.~\ref{n4bures},
 with the use of MATHEMATICA.
We obtained (using the Gauss-Kronrod integration method 
with a working precision of twenty-one digits, rather
the machine precision of sixteen) the result $2342475135.00/ \pi^{2}$.
Now, it is most interesting to note (particularly, in light of our
results for $C_{n}$, $n<5$, that
\begin{equation} \label{C5}
2,342,475,135 = 21  \Pi_{i=2}^{9} P_{i} = 21 (3 \cdot 5 \cdot 7 \cdot 11
\cdot 13 \cdot 17 \cdot 19 \cdot 23),
\end{equation}
where $P_{i}$ denotes the $i$-th prime number (taking the sequence of primes
to be $2, 3, 5, \ldots$).
Thus, we have acquired strong evidence that, in fact,
$C_{5} = 2342475135/ \pi^{2}$. (As a simple exercise, we looked at the
one hundred thousand
 consecutive integers containing 2,342,475,135 as their midpoint,
and computed all their prime decompositions. All the others had at
least one prime factor greater than 23.)
The seemingly independent factor of 21 in (\ref{C5}) will also apparently
be found below in the (odd) case $n=7$.

\subsubsection{Prime factorials}

 In \cite{caldwell}, the product of
the primes less than or equal to $p$ is denoted $p \#$.
(The issue there, as in several of the works cited there, was to test
$p \# \pm 1$ for primality (cf. \cite{shor}).)
Let us point out a 1952 article \cite{duarte}, entitled ``Tables of logarithms
of the prime factorials from 2 to 10007'' (a synopsis of which can be
found in Mathematical Reviews 16, 112f). It is noted that by the prime
number theorem \cite{primenumbertheorem} the ratio of the 
sum of the logarithms of the primes from 2 to $p$ to $p$ itself
approaches 1 as $p \rightarrow \infty$. 
So, if it eventuates that the general formula for $C_{n}$, at least for
odd $n$, contains a term of the form $p(n) \#$, 
where $p(n)$ is a prime as a function of $n$, which grows indefinitely
large with $n$ itself, then it should be possible to asymptotically
replace $p(n) \#$ by $e^{p(n)}$.
(``A version of the prime number theorem states that the product of the primes
less than $x$ is asymptotically $e^{x}$ [citing 
the well-known treatise \cite[Theorem 434]{hardy}],
but the error term is notoriously large, so it is probably unrealistic to
expect to be able to compute far enough to get within the necessary epsilon''
\cite{guy}.)
We also observe that with the use of Wilson's theorem \cite{wilson},
$(p-1) ! \equiv  -1 \mbox{mod} {p}$, one could express $p \#$ in
 terms of the (more) standard
factorial function $p!$. (In \cite{mat}, the primes are {\it defined} in terms
of factorials.)

Let us point out that Ellinas and Floratos have recently
studied the ``prime decomposition'' of an $n \times n$ density matrix into
 a sum of separable density matrices with dimensions determined by
the coprime factors of $n$ \cite{ellinas} (cf. \cite{number}).

\subsection{Preliminary investigations of the Hall constant for $n>5$, with the use
of quasi-Monte Carlo integration}

\subsubsection{$n=6$} \label{subn6}

Of course, as the dimensionality of the $n$-state quantum systems increases
(that is, $n$ itself increases), the numerical integrations required to
sufficiently narrow estimates of the corresponding Hall constant become
increasingly more difficult, and it is hard to judge what is precisely
the optimum numerical/programming strategy to employ.
Following the methodology outlined
 in
sec.~\ref{n4bures}, based on the ordering of the eigenvalues, MATHEMATICA
 did yield (using the standard default options)
an estimate of $C_{6}$, representable in the form,
 $1.4616286 \cdot 10^{16} / \pi^{3}$ (although diagnostics as to
inadequate precision were issued during the course of the computation). This
 result, coupled with our observation
of the pattern of $C_{n}$ for $n < 6$, might lead us to speculate that
the numerator of $C_{6}$ is either the seventeen-digit number,
$14,616,907,579,654,144 = 2^{41} \cdot 17^{2} \cdot 23$
or $14,623,504,649,420,800 = 2^{42} \cdot 5^{2} \cdot 7 \cdot 19$,
with the  denominators, in both cases,  being $\pi^{3}$.

When we  employed the  quasi-Monte Carlo 
(Halton-Hammersley-Wozniakowski) procedure \cite[chap. 3]{hammersley}
\cite{saar}
of MATHEMATICA,
 to numerically
integrate over a hypercube of volume $\pi^{5}$
(corresponding, thus, now to no particular distinguished ordering of the
six  eigenvalues), we obtained (with {\it no} diagnostics at all
being generated in two separate analyses --- having set maxima of ten and
fifty million sample points)
 a result of the form
$1.536355674 \cdot 10^{16} / \pi^{3}$. To a very high accuracy, this numerator
can be approximated by
$15,363,556,773,986,304 = 2^{28} \cdot 3^{4} \cdot 13^{2} \cdot 37 \cdot 113$
for the numerator of the presumptive value of $C_{6}$ (with denominator, again,
$\pi^{3}$).
 Sloan and Wo\'zniakowski have noted that recently
``Quasi-Monte Carlo algorithms have been successfully used for multivariate
integration of high dimension $d$, and were significantly more efficient than
Monte Carlo algorithms'' \cite{sloan}.
However, in comparing the two sets of results here for $n=6$ it
 is very important to bear in mind that MATHEMATICA sets its precision and
accuracy objectives much lower when Monte Carlo procedures are employed
\cite{saar}.
(Of course, these default values can be reset, but in the preliminary
 analyses reported
here, they have not been, though we intend to do so in future studies.)

\subsubsection{$n=7$} \label{subsecn7}

The quasi-Monte Carlo MATHEMATICA procedure produced an estimate 
(with no accompanying diagnostics, having set a maximum of four million
sample points) of
$C_{7}$ as $2.4811899
 \cdot 10^{24} / \pi^{3}$. 
The numerator of this fraction (cf. (\ref{C5}))
could be approximated  (to a relative error of less than three-tenths of
 one percent)
 by  the product of 21 and the (eighteen) consecutive primes
from 2 to 61 (in the notation of \cite{caldwell}, this is $61 \#$), that is,
$C_{7} = 21 \Pi_{i=1}^{18} P_{i} / \pi^{3}$.
However, this degree of accuracy does not at all seem satisfactory, 
particularly in light of the proximate results for $n=6$ and 8, though
odd $n$'s appear to present greater computational challenges.

\subsubsection{$n=8$}

The quasi-Monte Carlo procedure estimated
 $C_{8}$ as $4.1836777028  \cdot 10^{35} / \pi^{4}$ (though, unlike
the $n=6,7$ cases, a failure to converge was reported  ---  that is, 
with the preassigned use of at most
 ten million sample points). Nonetheless, this outcome 
can be fit to a very
high accuracy (less than one part in one hundred million) by taking
$C_{8}$ to be
$2^{89} \cdot 3^{2}  \cdot 5  \cdot 13^{2} \cdot 31 \cdot 47 \cdot 61 / 
 \pi^{4}$.

\subsection{$n \geq 9$}

For the case $n=9$, convergence was not obtained with the use of
six million sample points. The result given was 
$7.631832917  \cdot 10^{47} / \pi^{4}$.
We have also made tentative attempts, using the quasi-Monte Carlo procedure
again, to estimate $C_{n}$ for $n =10$ and 11 (but it appears that considerable investment of computer resources is  needed for sufficiently
satisfactory answers). For $n=10$, based on a maximum
of two million sample points, convergence was not obtained and the
result $6.334733996
 \cdot 10^{62} / \pi^{5}$ reported. (No related decomposition
was immediately apparent.) For $n=11$ and 12, using a maximum of one million
sample points in both cases, the results $3.254198489 \cdot 10^{78} /
\pi^{5}$ and $5.218327334 \cdot 10^{97}/ \pi^{6}$
were gotten
(without convergence or any obvious associated simple
prime decompositions, however).
Our impression is that the computations are considerably
more difficult for the odd values of $n$ than for the even, which may be
some reflection of the simpler formulas displayed above for even $n$
(cf. \cite{chen,cvij}).
(For $n>7$, the MATHEMATICA compiler was unable to handle the small numbers
appearing in the calculation, and then proceeded with the use of the
uncompiled evaluation, leading to slower running times \cite{saar}.)
Of course, there exists a wide range of 
possible approaches to numerical integration problems of this kind,
including, certainly, the use of alternative programming languages,
in particular, FORTRAN. The trade-offs
between these various options need to be assessed.

As a reference point, against which one can attempt to
compare the (reciprocals of the)
 several values of $C_{n}$ above, let us recall that the area of an
$n$-sphere of radius $r$ is given by $2 \pi^{n/2} r^{n-1} / \Gamma(n/2)$
\cite{fusaro}. Also of similar interest is Euler's formula
\begin{equation} \label{riemann}
\zeta(2 n) = {(-1)^{n-1} \over 2 (2 n)!} (2 \pi)^{2 n} B_{2 n},
 \quad n =1,2,3,\ldots
\end{equation}
where $\zeta(s)$ is the Riemann zeta function, $\sum_{n \geq 1} n^{-s}$,
 and $B_{n}$ is the
(necessarily rational) $n$-th Bernoulli number
 \cite[Vol. I, pp. 75 and 211]{terras}.
(The values of $\zeta(n)$ for positive odd integers $n$, however, have not been
expressed in such a simple form \cite[Vol. III, p. 1695]{ito}. Infinite
series representations are, in fact, reported in \cite{chen,cvij}.)
Initial attempts to find an explanatory formula for the sequence of 
(integral) numerators of $C_{n}$, using on-line programs
of C. Krattenthaler (``Rate'') \cite[App. A]{advanced} and of N. Sloane 
(``superseeker'') \cite{sloane}, did not succeed.
However, we did 
eventually find a somewhat intriguing  connection (further
buttressed by some related analyses, discussed below in sec.~\ref{further})
between the results here and sequence A035077 of \cite{sloane}, which gives the
denominators of partial sums of  $B_{2 n}$.
The numerator of: (1) $C_{2}$ is (rather trivially) twice
 the first entry in this
sequence; (2)  $C_{3}$ is one-half the fourth entry; (3) $C_{4}$
 is $2^{10} = 1024$
times the fourth entry; and (4) 
  $C_{5}$ is 63 times the thirteenth entry of A035077. (We note that 
Donaldson has found simple proofs of various formulas for {\it symplectic}
volumes involving Bernoulli numbers \cite{donaldson,witten,jeffrey}.)

\subsection{Average von Neumann entropy of $n$-state quantum systems with
respect to Bures prior probability distributions}
 \label{ENTROPY}

As one application of these computations of the Hall constants ($C_{n}$),
 let us note that with respect to the Bures
probability distribution (\ref{n2})
the average von Neumann entropy,
$-\mbox{Tr} \rho \log{\rho}$, is exactly
$2 \log{2} - 7/6 \approx .219628$ nats for the two-state systems and, 
now using numerical integration, 
.507937 nats for the three-state systems (cf. \cite{page}). (Since we
employ the natural logarithm here, the  unit of information is the
{\it nat}, which is equivalent to $1/ \log{2} \approx 1.4427$ {\it bits}.) 
This latter
result, to a high degree of precision --- that is, to ten significant places,
can, in fact, be written as $3 \log{3} - 3917/1405$.
We  also computed
the  Bures average entropy  for the four-state quantum systems, obtaining
.751771 nats. To eight significant places, this can be written
as $4 \log{4} - 32135/6704$. Similarly, for $n=5$, using our knowledge
of $C_{5}$, we obtain an average entropy of .954103 nats.
This is closely approximated (to at least nine places,
according to our calculations) by $5 \log{5} - 40045/5648$.

\subsection{Auxiliary analyses of variations of Hall integrals and 
the role of Bernouilli numbers} \label{further}

Since it appeared to be quite challenging to determine the Hall constants
($C_{n}$) for $n > 5$, we thought that it might be revealing 
(possibly helpful in deriving a general formula for arbitrary $n$), as well as
being of independent interest, 
to investigate
more tractable variations. To do so, we replaced
the exponent two in (\ref{hallformula}) and 
(\ref{hallconstant})  by either one (corresponding to {\it real} quantum
systems) or four (for quaternionic quantum systems) \cite{mehta}, and other
positive integer values of less immediate {\it physical} interest, as well.

For the $n=3$ (spin-1) quaternionic case (that is, using an exponent
of four), the counterpart of the previously derived 
(sec.~\ref{proximate}) Hall constant $35 / \pi $
is $1616615/ {226 \pi} = 5 \cdot 7  \cdot 11 \cdot 13 
\cdot 17 \cdot 19 / 226 \pi$. (The numerator here is one-half the tenth member of
the sequence A035077, comprised of
 denominators of partial sums of the Bernoulli 
numbers $B_{2 n}$ \cite{sloane}.) When we use an exponent of six,
 the result is $ 100280245065 /88252 \pi = 3 \cdot 
5 \cdot 7 \cdot 11 \cdot 13 \cdot 17 \cdot 19 \cdot 23 \cdot 29 \cdot
 31 / 88252 \pi$. (This numerator is precisely the seventeenth member of A035077.)

If we use an exponent of one (corresponding to the case of 
real quantum systems), the result is $1 /4 \pi$,
 and if we employ an exponent of
three, the normalization constant is 
$ 105 / 128 \pi = 3 \cdot 5 \cdot 7 / 128 \pi$. (The numerator here is
the fifth member of A035077.) For an exponent of five, we have 
$15015 / 8192 \pi = 3 \cdot 5 \cdot 7 \cdot 11 \cdot 13 /  8192 \pi$, 
the numerator being three times the seventh entry of Sloane's sequence.
For an exponent of seven, the result was the
eleventh member of the sequence A035077, that is, 969969, divided by
$262144 \pi$.

When an inquiry was made of 
Neil Sloane as to whether to his knowledge there were any published
discussions of this sequence, he replied ``No, I was just
looking at various sequences of important rationals, and thought that the
pair A035078/A035077 should be in the database''. (However, he later pointed 
out that the von Staudt-Clausen Theorem \cite[p. 10]{rademacher} was 
relevant to questions involving sums of Bernoulli numbers.) For the case 
$n =3$, MATHEMATICA rejected our
efforts to compute any further  exact integrals
 having integer exponents greater than seven. (It would appear that the use
of an exponent of eight would be associated with the {\it octonionic} 
quantum systems \cite{deleo}.)

For the analogous set of variations with $n=2$, use of odd exponents in
(\ref{hallformula}) and (\ref{hallconstant}) lead to divergent results.
For an exponent of two, the result is $2 \pi$, for four, $8 / 3 \pi $,
for six, $16 / 5 \pi $, for eight, $128 / 35 \pi $, \ldots.

For comparable scenarios based on $n=4$, we were unable to proceed with
exact integrations.
Our 
numerical 
computation of the analog of the Hall constant employing an exponent of
unity (the real quantum case), yielded $ \pi^{2} / .079271$. But we were
unable to determine if this result bore any relation to the sequence A035077.

\section{{\it QUASI}-BURES PROBABILITY DISTRIBUTIONS OVER THE $n \times n$
DENSITY MATRICES} \label{QUASI}

In line with the work reported in \cite{kratt}, it would be of interest
to obtain formulas for the averages over the eight-dimensional
convex set of $3 \times 3$ density matrices ($\rho$)
with respect to the Bures prior probability 
distribution (\ref{new}) of the $m$-fold tensor products of $\rho$.
As $m \rightarrow \infty$, the relative 
 entropy of these products with respect to the averaged
$3^{m} \times 3^{m}$ density matrix gives us the (Bures) asymptotic redundancy
for the universal quantum coding of {\it three}-state systems.

\subsection{The quasi-Bures case $n=2$}

In \cite{kratt} and further yet unreported 
work, the Bures prior probability density (\ref{n2})
 for the two-state
systems was found to closely resemble the (what we term
``quasi-Bures'') probability density,
\begin{equation} \label{wsquiggle}
\mbox{d} p_{quasi:2}(\theta,\alpha,\beta) = .226231 (\tan^{\sec{\theta}}{\theta / 2}) \cos{\theta}
 \cot{\theta} \sin{\beta} \mbox{d} \theta \mbox{d} \alpha \mbox{d} \beta,
\qquad 0 \leq \theta \leq {\pi \over 2}, \quad 0 \leq \alpha
\leq 2 \pi, \quad 0 \leq \beta \leq \pi
\end{equation}
which yields, it appears,
 {\it both} the asymptotic minimax and maximin redundancies
(as ``Jeffreys' prior'' \cite{kass} does classically \cite{cb1,cb2}).
This common value, if one ignores the error
term, as appears to be legitimate,
 is $(3 \log{m}) / 2 - 1.77062$, while the Bures
probability distribution (\ref{n2}) has been shown to
be, incorporating the error term, associated with an asymptotic redundancy of
$(3 \log{m}) /2 - 1.77421 + O(1/\sqrt{m})$ \cite[p. 29]{kratt}.
In general, for any probability distribution
 $w(\theta), \theta \in [0, \pi/2]$,
the asymptotic redundancy for the two-state quantum systems takes the form,
\begin{equation} \label{redundancy}
{3 \over 2} \log{{m \over 2 \pi}} -{1 \over 2} -2 \log{\sin{\theta}} +
2 (\sec{\theta}) \log{\tan{{\theta \over 2}}} -\log{w(\theta)} + o(1).
\end{equation}
Standard variational arguments can, then,  be used to show (ignoring the
error term, the legitimacy of which seems plausible, but has
not yet been rigorously justified) that the particular $w(\theta)$ yielding
 both the maximin and minimax
redundancies is simply proportional to the $\theta$-dependent part of
$p_{quasi:2}(\theta,\alpha,\beta)$, indicated in (\ref{wsquiggle}).

The reciprocal of the corresponding ``Morozova-Chentsov function'' $c(x,y)$
\cite{petzsudar,petzlaa} for (\ref{wsquiggle}) is the exponential 
or identric mean \cite[eq. (1.3)]{qi} of $x$ and $y$, 
\begin{equation} \label{qi13}
I(x,y) = e^{-1} (x^{x}/y^{y})^{{1 \over x-y}},\quad x \ne y
\end{equation}
($I(x,x)= x$),
while
for the Bures (minimal monotone) metric, it is the (more commonly
encountered) arithmetic mean $(x+y)/2$.
The associated operator monotone functions 
\cite{petzsudar,petzlaa} are $f(t) =(1+t)/2$
for the Bures metric, and $f(t) = t^{t/(t-1)}/e$, for the metric giving
(\ref{wsquiggle}). (The Morozova-Chentsov functions fulfill the
relation $c(x,y) =1/yf(x/y)$.)
Perhaps the {\it  exponential}
 mean arises in this context because the von Neumann
entropy is the {\it logarithmic} relative entropy, and of
course the exponential and logarithmic functions are inverses of one another.
This leads us to speculate that if one were to employ, following
\cite{lesniewski}, the ``quadratic
relative entropy'' or the ``Bures relative entropy'' instead, then,
in the parallel universal coding context, the minimax/maximin would be 
achieved  by the means corresponding to the
new forms of  inverse functions. While the logarithmic relative entropy
is based on the operator convex function, $g(t)=-\log{t}$, the quadratic form 
relies upon $g(t) = (t-1)^{2}$ and the Bures form on $g(t) = (t-1)^2/(t+1)$
\cite{lesniewski}. Here, $f(t) = (t-1)^{2}/(g(t) + t g(1/t))$.

\begin{figure}
\centerline{\psfig{figure=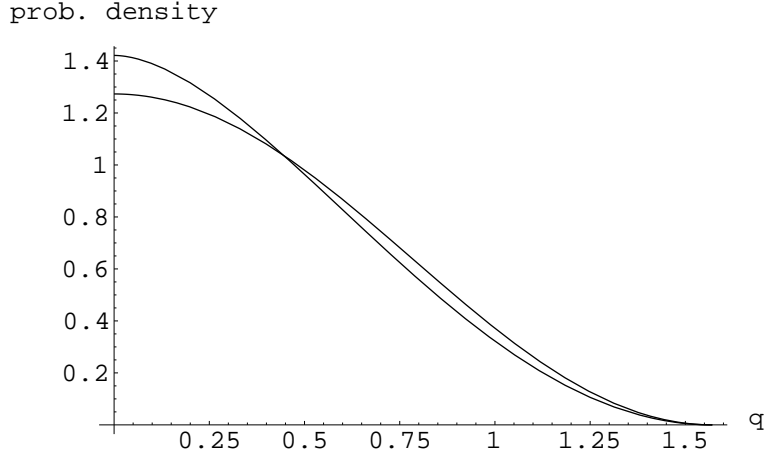}}
\caption{One-dimensional marginals of the ($n=2$)
 Bures probability distribution
(\ref{n2}) and the quasi-Bures distribution 
(\ref{wsquiggle}) (assuming greater values  for $\theta \in [0,.443978]$),
which
yields both the asymptotic minimax and maximin redundancies for the universal
quantum coding of two-state systems}
\label{comparison}
\end{figure}

In Fig.~\ref{comparison}, we jointly
 display the univariate marginal probability
distributions of   (\ref{n2}) and (\ref{wsquiggle}), revealing
that they closely resemble one another, with the quasi-Bures distribution
assigning relatively greater probability to the
states more pure in character ($\theta \leq .443978$).

\subsection{The quasi-Bures case $n=3$}

\begin{figure}
\centerline{\psfig{figure=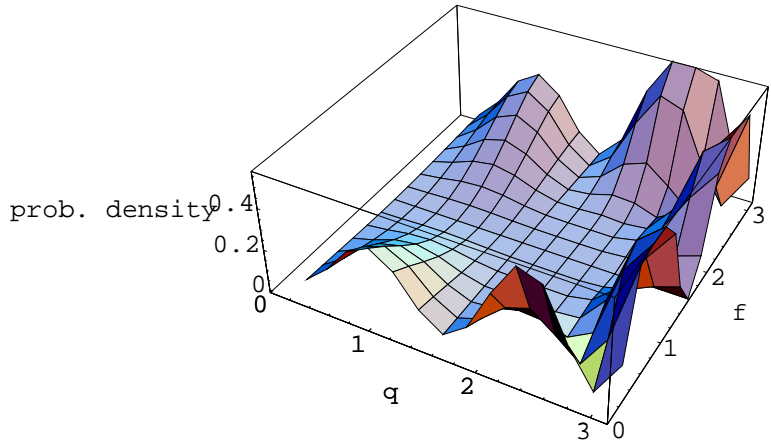}}
\caption{Bivariate marginal of the quasi-Bures probability distribution
(\ref{quasiBures})  over the
variables $\theta$ and $\phi$, for three-state systems}
\label{exponentialspin1}
\end{figure}

\begin{figure}
\centerline{\psfig{figure=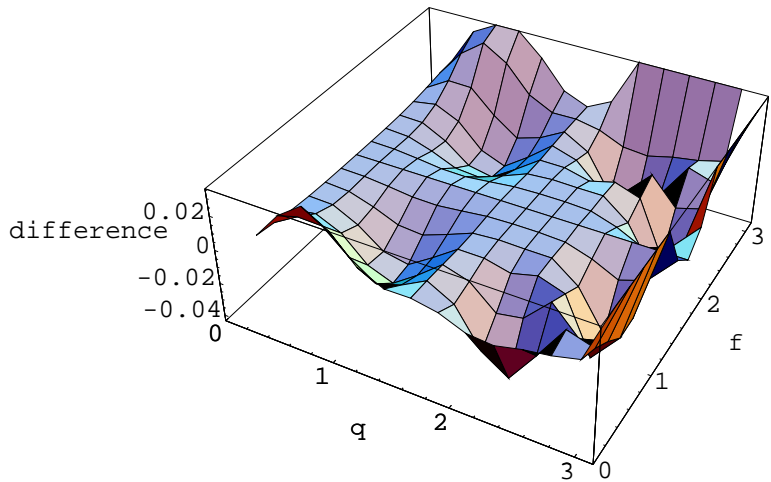}}
\caption{Difference obtained by subtracting the 
three-state Bures bivariate marginal probability density
in Fig.~\ref{twodim} from the quasi-Bures one,
 associated with the exponential/identric mean, displayed
in Fig.~\ref{exponentialspin1}}
\label{plotdifference}
\end{figure}

The three-state counterpart of (\ref{wsquiggle}) (that is, the probability
distribution associated with the exponential/identric mean, rather than 
the arithmetic mean, as for such Bures distributions) is 
\begin{equation} \label{quasiBures}
\mbox{d} p_{quasi:3}(\theta,\phi,\alpha,\beta,\gamma,\kappa,a,b) =
.000063495 u ({\tan^{(1 + \sec{\phi})}}{\phi / 2})
 {\csc^{4}{{\theta \over 2}}}
{\csc^{6}{{\phi \over 2}}} 
\end{equation}
\begin{displaymath}
 (\cos{{\phi \over 2}} \tan{{\theta \over 2}})^{{16 
{\cos^{2}{{\phi \over 2}}} {\sin^{2}{{\theta \over 2}}} \over {v+w}}}
(\sin{{\phi \over 2}} \tan{{\theta \over 2}})^{2 -{8 (1 + \cos{\theta})
 \over {w -v}}} \mbox{d} \theta \mbox {d} \phi \mbox{d} \widetilde{\Omega}_{3},
\end{displaymath}
where $u$ is given in (\ref{newvariable}),
$\mbox{d} \widetilde{\Omega}_{3}$ in (\ref{conditionalhaar}) and
\begin{equation}
v = 2 + 6 \cos{\theta}, \quad w = \cos{(\theta-\phi)} -2 \cos{\phi}
+\cos{(\theta +\phi)}.
\end{equation}
 The corresponding two-dimensional marginal probability distribution
over the variables $\theta$ and $\phi$ is exhibited in 
Fig.~\ref{exponentialspin1}.
As would be anticipated from Fig.~\ref{comparison}, this figure
 closely
 resembles Fig.~\ref{twodim}.
 In Fig.~\ref{plotdifference},
we show the result obtained by subtracting the  bivariate marginal 
Bures probability
distribution shown  in Fig.~\ref{twodim} from its  quasi-Bures counterpart
 in Fig.~\ref{exponentialspin1}.

\subsection{Quasi-Bures counterparts of Hall (Bures) normalization constants}

If we replace the term $d_{j} + d_{k}$, occurring in the denominator of
the expression (\ref{hallconstant}), by (twice) the exponential/identric mean
(\ref{qi13}) of $d_{j}$ and $d_{k}$, that is,
 $I(d_{i},d_{j})$, the resultant expression becomes a formula
for $\tilde{p}_{quasi:n}(d_{1},\ldots,d_{n})$, now interpreting $C_{n}$ to
be the normalization constant for the 
corresponding quasi-Bures probability distribution. For $n=2$ then
(taking $\theta \in [0,\pi]$ and not $[0,\pi/2]$ as in 
\cite{boya}),  rather
than $2 / \pi \approx .63662$, we obtain .769427, and for $n=3$, instead of
$35 / \pi \approx .089759$, we find .138681.

\section{CONCLUDING REMARKS}

We hypothesize that the (full eight-dimensional) 
quasi-Bures probability distribution (\ref{quasiBures})
associated with Fig.~\ref{exponentialspin1} will
 furnish the common asymptotic (minimax and maximin)
redundancies for universal quantum
coding
in that higher-dimensional setting, paralleling the 
result (not yet fully formally demonstrated, however) for the
three-dimensional convex set of two-state systems (cf. \cite{jozsa}).
In this regard, 
 it might prove computationally convenient, as a heuristic device,
 to replace the
quasi-Bures probablity distributions by their (closely approximating) Bures
analogues, since certain exact (symbolic) integrations are achievable 
(at least, for the cases $n = 2, 3$) with
the Bures distributions, but apparently not with the quasi-Bures ones,
for which numerical methods seem to be necessary.

Although a parameterization of SU(4) is not relevant, as already
noted, to the computation
of the Hall constant $C_{4}$, that is, $71680 / \pi^{2}$, and to that of
the corresponding average (von Neumann) entropy (sec.~\ref{ENTROPY}),
 it would be essential in investigating the
universal coding of four-state quantum systems, since the tensor products of
$4 \times 4$ density matrices  have to be calculated and it would,
therefore, 
be necessary to implement formula (\ref{one}). In a personal communication,
M. Byrd has indicated that he has undertaken the (challenging) task of
developing such an 
(Euler angle) parameterization of $SU(4)$.

Such a parameterization of $SU(4)$ --- in conjunction with the knowledge,
acquired here, of $C_{4}$ ---  might also prove of value
 in estimating the volume
of $2 \times 2$ separable quantum  states \cite{zhsl,zyczrecent}
 and lead to numerically
more stable results than those
 reported in \cite[Table I]{slaterjpa2}, since the ``over-parameterizations''
of the unitary matrices used there
could then be avoided, due to the ``dropping out'' 
(as pointed out for $n=2,3$ immediately after (\ref{ranges}) above) of
certain Euler angles in the formation of the product (\ref{one}).
 We also note
that in \cite{slaterjpa2} (cf. \cite{slaterduan}),
 the Bures (minimal monotone) metric was found
to yield higher {\it a priori} probabilities of entanglement than other
monotone metrics (in particular, the Kubo-Mori-Bogoliubov and maximal ones).
Presumably, even 
in any computationally improved form of analysis, this conclusion
would be unaltered.

We have also pursued a traditional (pseudo-random number) Monte Carlo
approach to estimating the Hall normalization constants  for $2 \leq n \leq
16$. However, the degrees of precision  attained were not satisfactory.
(For  discussions of the comparative
 computational complexities of the pseudo- and
quasi-Monte Carlo methods, see \cite{sloan,frank}.)

 We are presently attempting
to obtain a more precise estimate of the Hall 
constant $C_{6}$, in particular, using {\it non}-Monte Carlo
 (that is, adaptive) 
integration
methods.
Our best current estimate of $C_{6}$ 
 is $1.534836628 \cdot 10^{16} / \pi^{3}$.
The numerator can be very well approximated by $15348366279966720  = 
2^{33} \cdot 3 \cdot 5
\cdot 7^{2} \cdot 11 \cdot 13 \cdot 17$, which seems more satisfactory than
the results reported in sec.~\ref{subn6}, based on the quasi-Monte Carlo
procedure. Now, it is most interesting to note that this numerator
 is precisely
$2^{33} \cdot 7$
 times the ninth entry of the sequence A035077, we have repeatedly
referenced above. We are also compelled to observe that our educated
conjecture as to the numerator of $C_{7}$ (sec.~\ref{subsecn7}) 
is exactly forty-two times the
thirty-third entry --- which we had to compute ourselves, since Sloane's
published list does not extend this far --- of A035077.

\acknowledgments

I would like to express appreciation to the Institute for Theoretical Physics
for computational support in this research and to M. Byrd and K.
\.Zyczkowski each for a number of helpful communications, as well as to
C. Krattenthaler for his insightful analyses, and to M. J. W. Hall for 
pointing out to me an erroneous statement in an earlier version.
Also I thank J. Stopple for the reference to 
\cite{terras} and his interest in this work and to
 M. Choptuik for a discussion concerning the relative
merits of various numerical integration routines.

\end{document}